\documentclass[letterpaper, 11pt]{IEEEtran}
\pdfoutput=1
\usepackage[margin=1.0in]{geometry}
\renewenvironment{IEEEbiography}[1]
  {\IEEEbiographynophoto{#1}}
  {\endIEEEbiographynophoto}

\usepackage[utf8]{inputenc}
\usepackage{cite}
\usepackage{amsmath,amssymb,amsfonts}
\usepackage{algorithmic}
\usepackage[pdftex]{graphicx}
\usepackage{textcomp}
\usepackage{xcolor}
\usepackage{url}
\usepackage{booktabs}
\usepackage{multirow}
\usepackage{adjustbox}
\usepackage[caption=false,font=footnotesize]{subfig}
\usepackage{hyperref}

\begin{document} 
 
\title{Mixed-Reality Robotic Games: Design Guidelines for Effective Entertainment with Consumer Robots\\
}

\author{

\IEEEauthorblockN{
        F.~Gabriele~Prattic\`o,
        and~Fabrizio~Lamberti
}

\IEEEauthorblockA{
    Politecnico di Torino, Dipartimento di Automatica e Informatica, Torino, Italy
}
}%

\markboth{IEEE CONSUMER ELECTRONICS MAGAZINE,~Vol.~XX, No.~X, XXXX~XXX}%
{Shell \MakeLowercase{\textit{et al.}}: Bare Demo of IEEEtran.cls for IEEE Journals}

\maketitle

\begin{abstract}
In recent years, there has been an increasing interest in the use of robotic technology at home. A number of service robots appeared on the market, supporting customers in the execution of everyday tasks. Roughly at the same time, consumer-level robots started to be used also as toys or gaming companions. However, gaming possibilities provided by current off-the-shelf robotic products are generally quite limited, and this fact makes them quickly loose their attractiveness. A way that has been proven capable to boost robotic gaming and related devices consists in creating playful experiences in which physical and digital elements are combined together using Mixed Reality technologies. However, these  games differ significantly from digital- or physical-only experiences, and new design principles are required to support developers in their creative work. This papers addresses such need, by drafting a set of guidelines which summarize developments carried out by the research community and their findings. 
\end{abstract}

\section{Introduction}
\label{sec:intro}
In the form of commercial off-the-shelf (COTS) products, robots are getting ever more commonplace in our lives. Service robots (vacuum cleaners, lawnmowers, etc.) and toy robots, in particular, constitute today a key market for consumer electronics (CE) \cite{cemagrobot}. While technology is attaining maturity and the economic relevance of the sector is growing, industry and academy are dedicating significant efforts in improving Human-Robot Interaction (HRI) with the aim to encourage acceptance of this technology and stimulate adoption. 

One of the emerging trends pertaining to robotics and CE is their application to gaming. A great variety of game scenarios built onto robots exist already, from drone races to robotic pets and home toy mates. The increasing diffusion of AI could also make these systems more smart and flexible. 

Notwithstanding, as of today, the most common use of these robots is through teleoperation. The constrained application scenarios provided out-of-the-box and the frequently unsatisfactory HRI patterns often make these experiences quickly lose their attractiveness. Hence, giving CE robots originally meant for a specific application a ``new life'' is regarded as essential to stimulate the interest towards this new form of entertainment and, consequently, towards the market of consumer robotics. 

To cope with this challenge, several directions have been explored. One of the most promising approaches takes advantage of another fast-growing field in CE: Mixed Reality (MR). Nowadays, thanks to the advancements in electronics and the growing interest in consumer-grade augmented reality (AR), MR is becoming widespread \cite{cemagar1, cemagar2}. MR applications usually exploit AR displays to present digital contents to the users, but differ from pure AR in that contents are  ``contextualized'' in the real environment (Fig. \ref{fig:pirg}): the goal is, in fact, to blend the border of what is real and what is not. A concern about interaction in MR is thus about how to provide interfaces capable to preserve this illusion and the sense of presence. Nowadays, MR interaction is moving from point-click paradigms to immersive interfaces based on body and hand gestures. In such a scenario, the intrinsically dualistic nature (physical and digital) of robots is a perfect match for the mix of these two ``big things'' in CE, as they largely complement each other.

\begin{figure}[t!] 
    \centering
        \subfloat[]{\includegraphics[width=0.48\columnwidth, height=3.0cm]{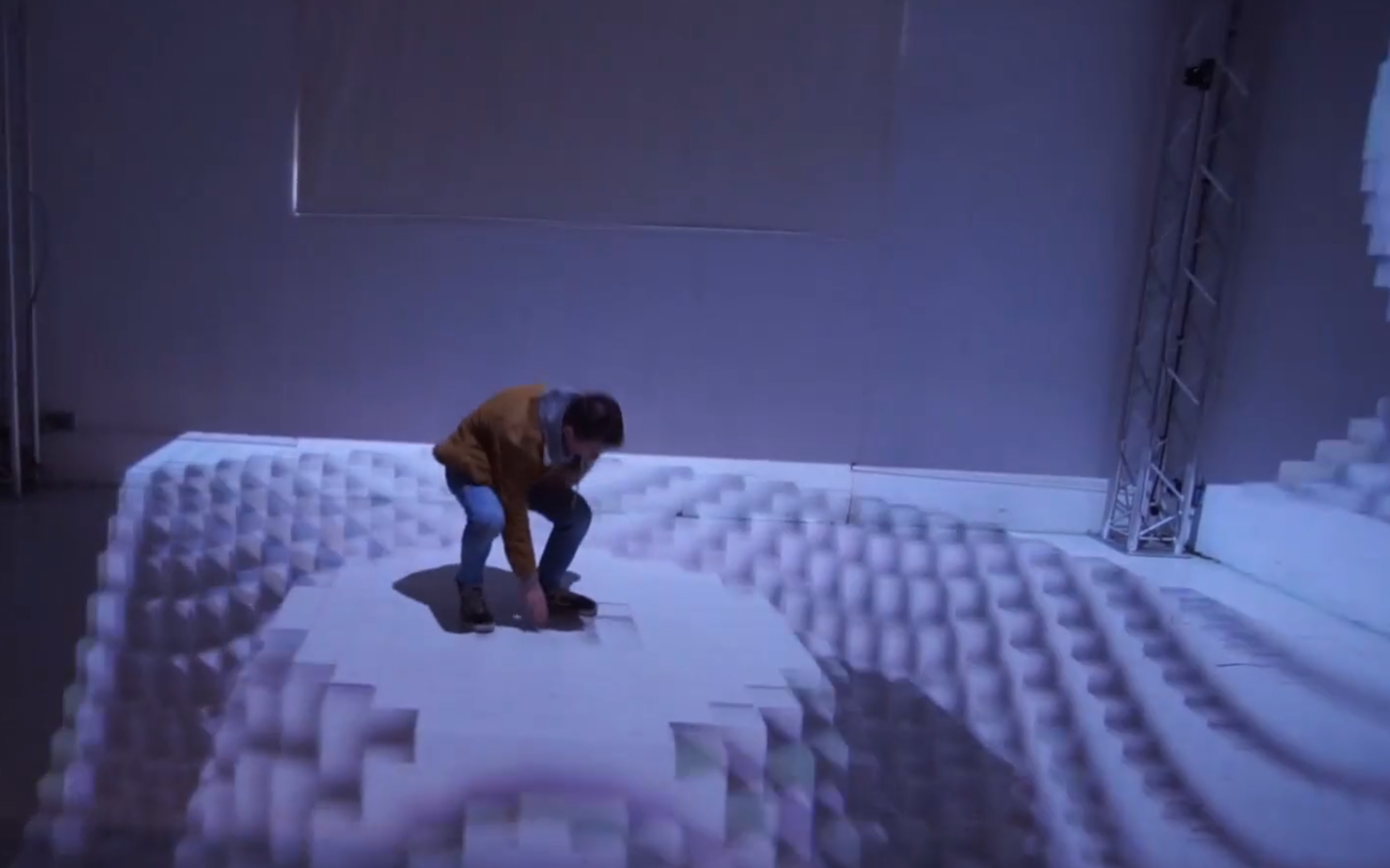}%
        \label{fig:pirg}}
    \hfil
        \subfloat[]{\includegraphics[width=0.48\columnwidth, height=3.0cm]{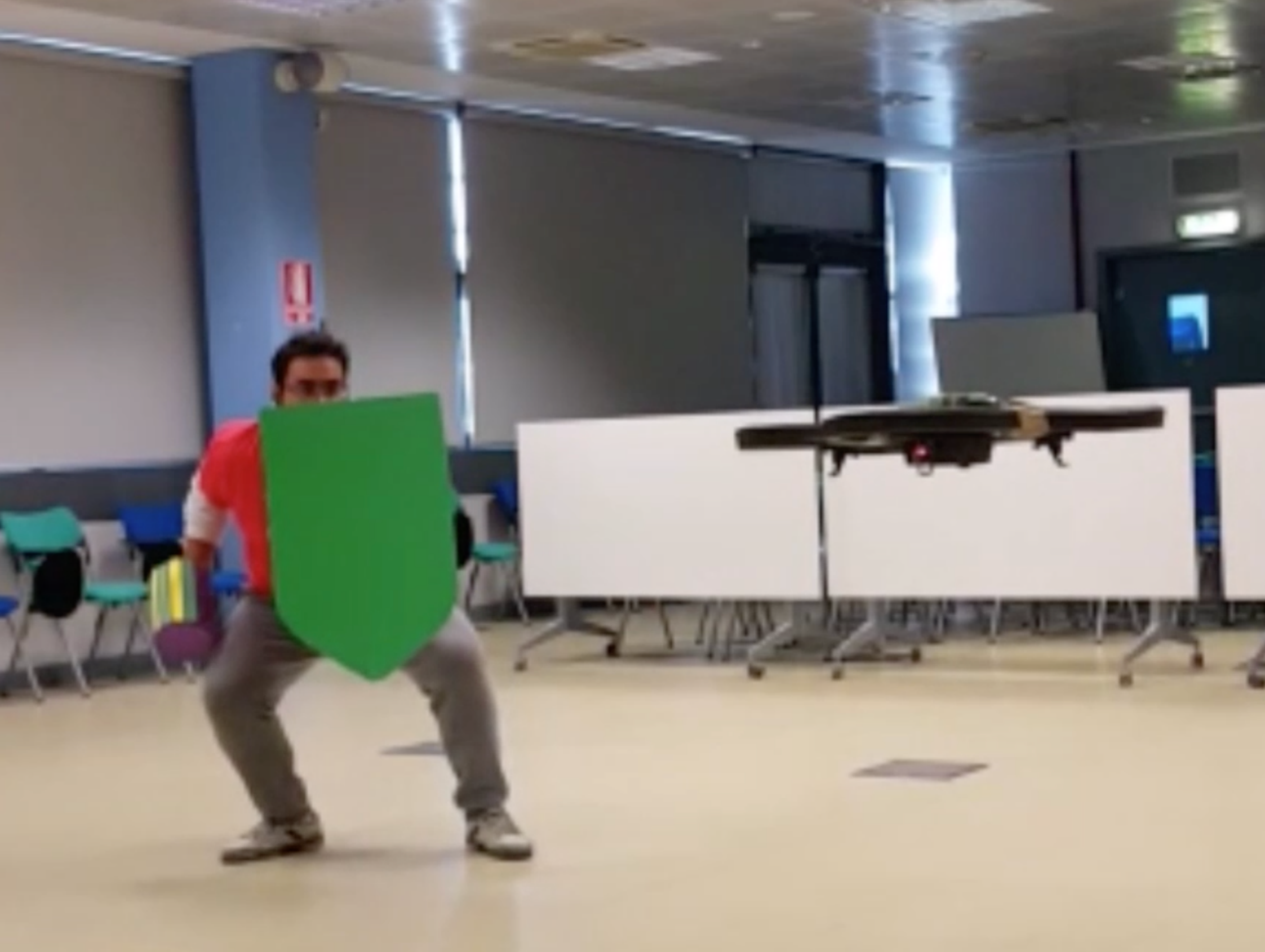}%
        \label{fig:MRRG}}
    \caption{Examples of a) Mixed Reality (MR) (\emph{Theoriz} -- CC BY 3.0) and b) Physically Interactive Robotic Game (PIRG) \cite{villani} experiences, where the digital and the physical components may be predominant, respectively: in a MR-based Robotic Game (MRRG), these two features are combined together.}
    \label{fig:pirg_vs_MRRG}
\end{figure}

Many studies were indeed performed in the last decade focusing on the development of games that include a mixture of physical and digital elements by leveraging MR, giving birth to the concept of ``phygital play'' \cite{phygital}. This form of interaction was investigated also in game scenarios involving robots, which led to the appearance of so-called Physically Interactive Robotic Games (PIRGs) \cite{pirg, villani}. While, generally speaking, in phygital play scenarios
the MR component is not strictly required, it is also true that in some cases such aspect can overshadow the physical one, which may be limited, e.g., to the use of natural interfaces (NIs). In PIRGs, instead, tight requirements are generally set on the physical part, due to the use of robotic elements; however, MR augmentation is not strictly necessary (Fig. \ref{fig:pirg_vs_MRRG}b). 

This work focuses on 
MR-based Robotic Games (MRRGs), where both robots and MR play a central role, thus setting new challenges for game developers. Despite common elements shared by the above applications, key differences between them could make the guidelines proposed for designing PIRGs \cite{pirg} and phygital play experiences not effective, or even counterproductive, when applied to MRRGs. Similar considerations apply to common tools and principles for the design of conventional games (digital or not) \cite{gdw}, which might need to be extended to deal with the specificities of MRRGs. 

Even though MRRGs have a great engagement and entertainment potential, it is not straightforward to extract it. It is not uncommon, to design and implement an MRRG and then find out that it is equally, or even more, fun to play without a robot or the MR component. A fair number of works addressed already the challenges associated with the creation of MRRGs from the technological point of view, by developing solutions for reusable and cost-effective setups based on many CE technologies \cite{pirone, robotable2, saro}. However, there is still a lack of structured assumptions and design principles to rely onto for building effective MRRGs: taking an existing MR game and trying to fit the robot to it, or viceversa, generally does not work as expected. 

By summarizing findings from research activities carried out at Politecnico di Torino in the last five years and combining them with results reported in the literature about HRI, this paper aims to provide a summary of design guidelines that should be considered for the making of engaging MRRGs, by also listing key technologies for building them.

\section{MRRG Definition and Challenges}
\label{sec:background}

\subsection{Definition of MRRGs}\label{subsec:def}

A MRRG is \emph{``a game in which at least one human being and one robot interact in a shared MR environment''}. No restrictions are set on the robot control scheme (autonomous or teleoperated). 

In principle, any programmable robot is eligible to be used in a MRRG. The most common in literature are CE robots (in particular toy and service robots), but it is certainly possible to design a MRRG leveraging, e.g., industrial robots (manipulators, mobile platforms, etc.). Indeed, there are some robots, with specific sensing and actuation capabilities, which shall be preferred (like mobile ones). Thus, it should not be surprising that a game designed for a given robot (or class of robots) is not effective when played, if feasible, with another one.

Similarly, the MR environment is intended in the broad sense, with digital, AR-based contents that can be delivered to the player(s) by employing any kind of setups (head-worn, hand-held, or spatial, i.e., using projection). Certainly, as per definition of MR, the digital contents must be ``aware'' of the physical environment, and respond to interactions.\\ 
Regarding target audience, works in the literature mostly focused on children, teenagers, and elderly people; there is also a large amount of studies that targeted people with specific diseases (Autism, Alzheimer, etc.) with so-called ``games with a purpose''. However, this age categorization could be biased by the fact that, as said, many works dealt with aspects pertaining technology rather than HRI or user experience (UX); thus, validations were often performed with subjects who did not match the actual end-users. Hence, for defining the target audience of MMRGs it is safer to apply classical Game Design (GD) principles \cite{gdw}, not expecting, e.g., that a game designed for a given user category provides the same fun to a different category. 

\subsection{Challenges in the Design of MRRGs}\label{subec:challenges}

Since these games appeared to have lots in common with PIRGs, the approach pursued in the studies conducted so far has been to build upon that solid basis, and apply the GD guidelines given for PIRGs \cite{pirg} (together with classical GD principles) to the creation of several MRRGs, evaluating the impact on the UX and on the robot role as perceived by the players. However, there are many challenges that are specific of MR-based games using robots.

\subsubsection{Identifying Challenges in MRRGs}\label{subsubsec:challenges}
To pinpoint key challenges, we conducted a literature review. We searched two databases, namely IEEE Xplore and the ACM Digital Library, since they include many works on robotics, Human-Machine Interaction (HMI) and MR. Moreover, sources like, e.g., the Journal of Social Robotics or Interaction Studies were added to consider also social aspects. The search string \emph{``robot* AND (gam* OR experience?) AND (mixed?reality OR augmented?reality OR mr OR ar))''} was used on titles. The cumulative number of unique works found was 38. Then, a broader query with the same string was performed on metadata, and the result set including more than hundred papers was refined by considering works published in the last 15 years having in the abstract terms like evaluation, comparison, guideline, UX, challenge, and assessment. This filtering allowed us to identify the most relevant works addressing MMRG-related aspects from the perspective of HRI \cite{HRIandMR}. We clustered them based on the tackled study dimensions and reported findings, and the most representative  ones were cited in the paper.

\subsubsection{Robot Irreplaceability Factor}\label{subsubsec:RIF}

Regardless of the particular technology used, there is a threat that any designer should be aware of while creating a MRRG which is truly specific for these experiences: \emph{how to promote and exploit the fact of having a real robot instead of a digital replica of it}.
 
\begin{figure}[t!] 
    \centering
        \subfloat[]{\includegraphics[width=0.48\columnwidth, height=3.0cm]{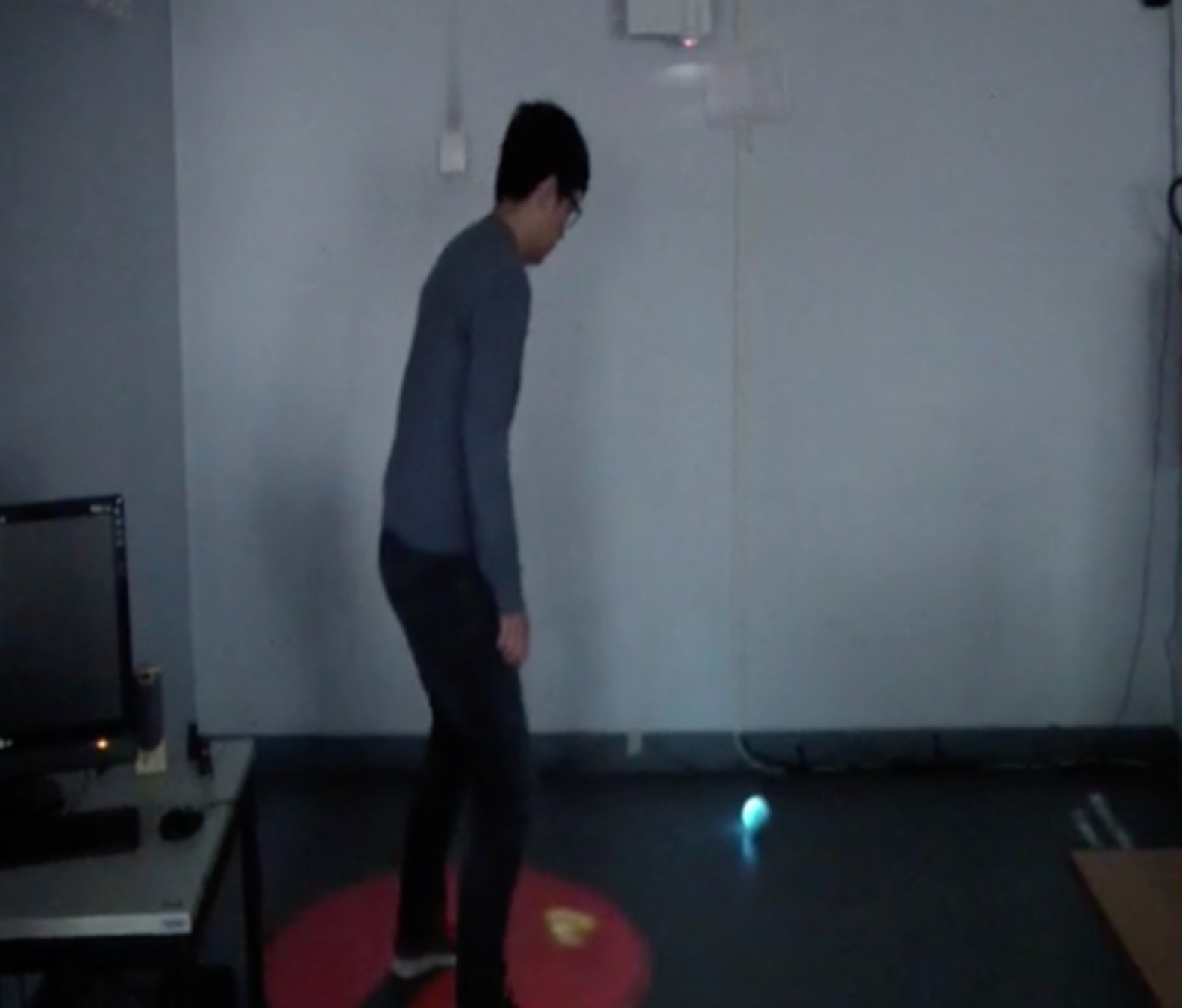}%
        \label{fig:physical}}
    \hfil
        \subfloat[]{\includegraphics[width=0.48\columnwidth, height=3.0cm]{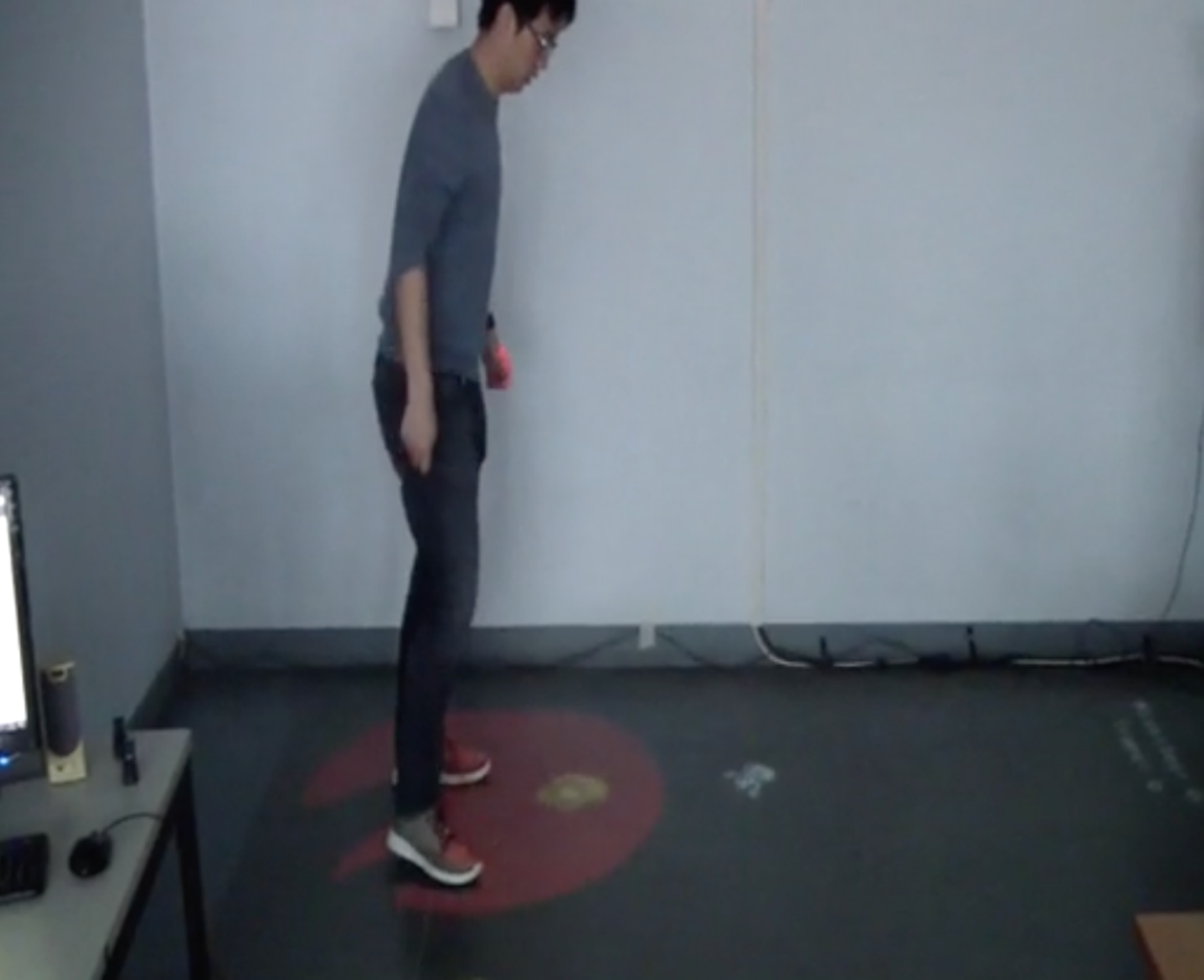}%
        \label{fig:virtual}}
    \caption{Replacing a) the physical robot of a MRRG \cite{junchao} (lighted ball) with b) a virtual replica (the projected ghost trying to catch the gold nugget).}
    \label{fig:physical_vs_virtual}
\end{figure}

Ironically, this subject has been actually covered in the literature from the opposite perspective. There are a decent amount of studies that focused on how to effectively replace the real (possibly robotic) elements barely altering  the UX. However, in MRRGs, replacing the robot with a virtual agent (VA) would immediately destroy the essence of the phygital play experience, collapsing the design in the broader category of MR-only games, without robots (Fig. \ref{fig:physical_vs_virtual}) \cite{junchao}. This situation is expected to worsen with advancements in MR technology, which will make the creation of realistic and believable holograms easily available at consumer level.

In our experience, the solution to this problem is, simplistically, to \emph{put the robot at the centre of the interaction}. This is actually not a real action to perform: it should be interpreted more as a good result of a well-designed MRRG, and could be regarded as an indicator of the fact that the physical robot is worth to be exploited in the game.

To refer to this unique challenge of MRRGs, we introduce a so-called \emph{Robot Irreplaceability Factor (RIF)}, which could be measured, e.g., on a 1--5 scale: 1) robot entirely replaceable by a VA;
2) VA has key advantages, but robot could be enjoyable;
3) neither robot nor VA have clear advantages over the other;
4) robot has key advantages, but a VA would suffice;
5) robot irreplaceable by VA.

In the following, we will present several RIF-driven GD guidelines which, in reviewed experiences, showed to be capable to boost this factor and, thus, should help to create more effective MMRGs (expressing expected impact on the above scale).

\subsection{Pivoting on the Robot Role}\label{subec:robot_role}

In principle, many strategies could be pursued to design a MRRG, like, e.g., devising a game concept, then searching for robots that could work for it. To lay out our guidelines we pinpointed in advance the robot role, since this choice shall empirically lead to more general design principles and was proven to be less prone to the creation of dependencies among the various design dimensions. In a MRRG, the robot can act matching the roles reported in the following (Fig. \ref{fig:roles}).
\begin{itemize}
    \item \emph{Adversary (ADV)}: the robot behaves as an opponent for the human player. The winning condition for the robot must be in contrast with the player's one.
    Although this is the most intuitive role for a robot in a MRRG, from the perspective of RIF it is the most though one. Evidences show that the RIF and the engagement of a game with ADV robot depend on the identification of a balance among the various game elements rather than on the adoption of specific principles, and are strictly related to the robot involved and its capabilities \cite{junchao}.

    \item \emph{Buddy (BD)}: the robot and the player have a shared winning condition; however, they could have different goals and follow different rules. One of the best ways to maximize the RIF in this case seems to provide the robot with skills complementary to those of the human which can be regarded as ``superhuman''. The optimal condition is obtained when these skills are enabled by native capabilities of the robot \cite{junchao, destefanis}: for instance, if a robot is equipped with a thermal camera, exploiting it in a game with a BD robot could increase the RIF more than ``emulating'' it with sintetic MR contents.

    \item \emph{Rule Keeper (RK)}: the robot is responsible to enforce game rules against the other agents (humans or robots). It is like some of the features of the game are transferred, better, embodied in the robot. There are many specializations of this role, such as, the referee or all-knowing wise-man \cite{baldo}, the game master \cite{dungeon_master}, etc. This is a very effective role in MRRGs, as it is suitable also for robots with limited capabilities (e.g., static robots), and can foster the perception of robots as intelligent agents.
\end{itemize}

It is worth noting that some roles fit specific MRRG applications better than others. For instance, in games designed to develop social  and interaction skills, it may be preferable to select the BD rather than the ADV role, with the aim to build an emotional bond between the human player and the robot. In games with training purposes, a good option would be the RK, as players seem to appreciate the feedback received from a robot better than that coming from a human instructor \cite{robo_train_feedback}.

\section{Design Guidelines}
\label{sec:designguidelines}
Below, we discussed the MRRG design dimensions that, based on current literature, could have an impact on the RIF. Other guidelines inherited from classical GD principles or PIRGs [5] possibly valid also for MRRGs (the ability of a robot to express emotions, the role of timing, the K.I.S.S. and the R\textsuperscript{3}, Reduce, Reuse and Recycle principles, etc.) but with no influence on the RIF, are not reported. 

\subsection{Player-Robot Interaction Pattern}\label{subec:p_r_int}

A first aspect to consider in designing MRRGs is the selection of a player-robot interaction pattern (IP). In conventional games, this is defined as \emph{``the structure of interaction between a player, the game system, and any other players''} \cite{gdw}. 

In digital games, two main IP categories are defined: competitive and cooperative. Each category can be further organized in several sub-patterns: player vs game (the most common in digital games), player vs player, multiple players as individual vs game, team competition, multilateral competition (like, e.g., in poker), etc.

\begin{figure}[t!] 
    \centering
        \subfloat[]{\includegraphics[width=0.48\columnwidth, height=3.0cm]{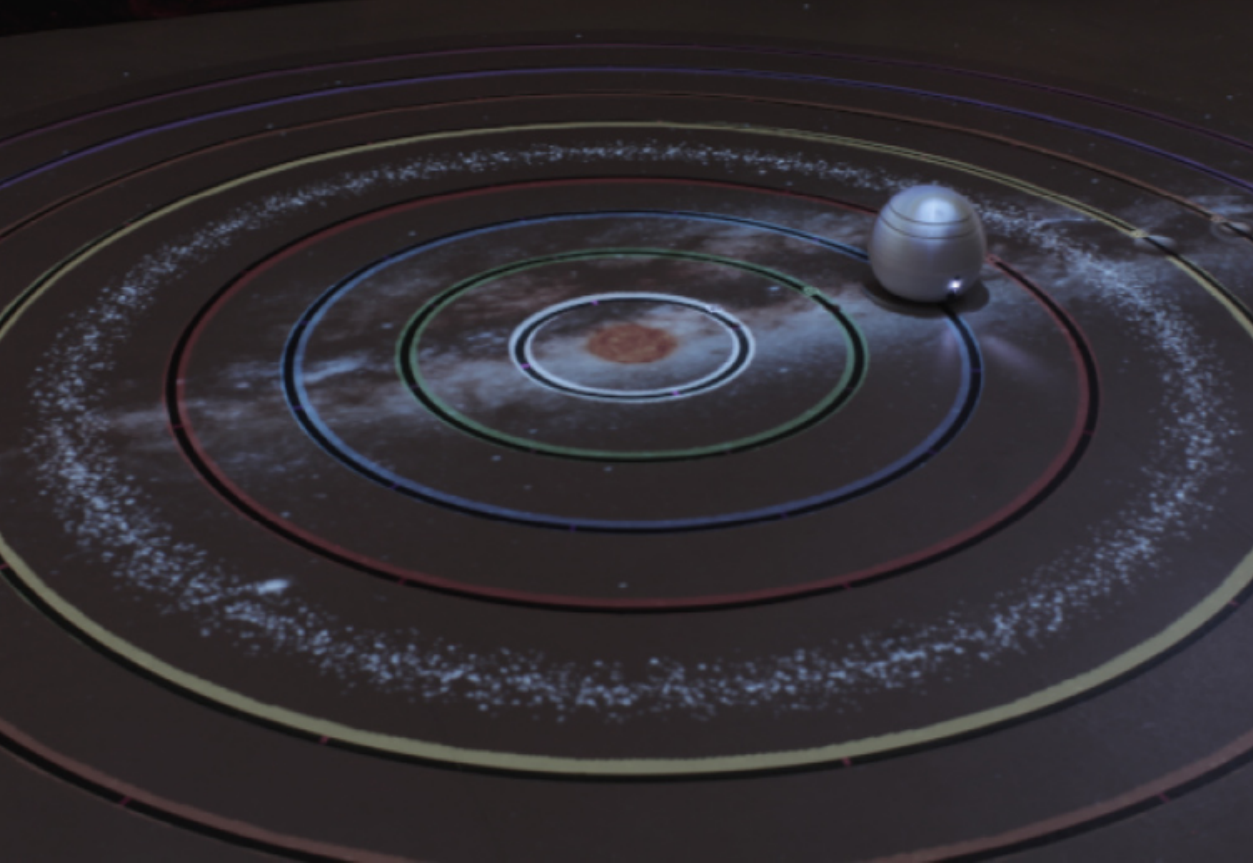}%
        \label{fig:solarsystem}}
    \hfil
        \subfloat[]{\includegraphics[width=0.48\columnwidth, height=3.0cm]{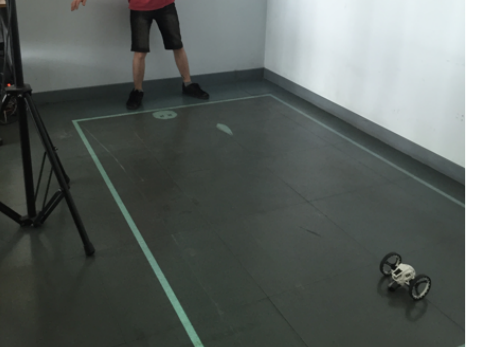}%
        \label{fig:bomberman}}
    \caption{Player-robot interaction patterns: a) rule keeper-based MRRG with educational purposes where the robot is in charge of asking questions on the Solar System and checking answers given by children with contactless cards; b) MRRG in which the robot is an adversary and shoots at the player \cite{saro}. }
    \label{fig:roles}
\end{figure}

It important to stress the difference between robot roles and the concept of IP. In fact, it is possible to design different games maintaining the same IP and changing the robot role. For instance, considering one player and one robot, it is possible to have an IP where the player is competing against the game system and the robot assumes either the role of BD or ADV (competitive players vs game). In both cases, the robot role would be considered as part of the game mechanics, without altering the resulting IP. Another (more complex) example of IP could be that of a game with a player and three robots, where each robot assumes a different role: a robot could be the game master (RK) enforcing the rules in a team formed by a player and a BD robot which compete against an ADV robot. 

The IP dimension should nonetheless be considered as tightly bonded to the game interaction modality: real-time (RT) or turn-based (TB). In RT games, players (and robots) interact simultaneously, whereas in TB games the interaction of the agents is regulated by turns (like, e.g., in chess).

A combination of IP and roles that, based on works like, e.g. \cite{baldo}, seems to provide the highest RIF (4/5) is represented by competitive multiplayer games involving human players (either singles or teamed) and a robot in the RK role. With this configuration, the modality to choose for maximizing the RIF is the TB one. It is, of course, possible to design a RT game with that configuration, but the RIF would be affected by the robot movement capabilities (Section \ref{subsec:movstyle}). The single-player version of this IP (one player vs the game) proved to be effective too, both with a robot as BD \cite{cellulo} and RK \cite{baldo}. The ADV role is the more delicate to fit in an effective IP, since the RIF is affected by many factors in this case (Sections \ref{subsec:mrc} and \ref{subsec:movstyle}). In other words, there is no preferred IP for this role.

\subsection{Robot Control}\label{subsec:control}
Among the possible design dimensions, the importance of the way the robot is controlled is generally underestimated. As said, the most common way of interacting with robots in games, at least in commercial solutions, is teleoperation (i.e., direct control). However, in MRRGs (and other applications as well), this control mode was proven to be the least effective way to foster the RIF compared, e.g., to autonomous behaviors (AI control) but also to approaches based, e.g., on Wizard of Oz (WoOz) techniques (in which the robot is directly controlled by another player who, however, is not considered as an agent in the IP, but just as the robot ``brain'').

\subsubsection{Considerations for ADV}
In particular, based on \cite{junchao} (where several ways to control the robot in different roles were analyzed by measuring the impact on player's perception), it is suggested to avoid the direct control of the robot in case it is playing the ADV role. For this role, it is suggested to empower the robot with an AI, or alternatively, use the WoOz approach. In the latter case, as demonstrated in \cite{bubusettete}, players perceive the robot as more intelligent, alive and lifelike if they are unaware that it is controlled by another human and, according to \cite{wozimpact}, they also change their behavior.

\subsubsection{Considerations for BD}
Direct control is acceptable, but it should restricted to limited periods of the gameplay or to a limited set of commands that can be triggered in addition to the AI- or WoOz-controlled ones \cite{destefanis}. For instance, the player could be provided with the ability to turn a given capability of the robot on and off (e.g., the thermal camera mentioned in a previous example), but the other behaviors should be controlled by an AI.  

\subsubsection{Considerations for RK}
The only control mode that makes sense for that role is the AI-based one (WoZ can be used for prototyping), as this role generally involves behaviors that can be fully automated without any support by the human \cite{oliveria_empathic}.

\subsection{Player Control and Interactions}\label{subsec:uint}
As it will be discussed in detail in Section \ref{subsec:mrc}, to interact with the digital elements of the game the player is provided with a digital counterpart of it (not strictly an avatar), which is not necessarily displayed during the game. Thus, the player interacts with the game physically, as well as controlling its digital counterpart. The various ways to implement this control have a role in maximizing the RIF.

Generally speaking, in MRRGs the use of conventional gaming devices, such as gamepads, joysticks, etc. is discouraged. The most effective control method appears to be represented by gestures and movements, in general, preferably performed using the entire body (or parts of it) \cite{villani, saro} and possibly combined with other NIs (like voice, gaze, etc.) \cite{junchao, baldo}. 
When the virtual counterpart matches exactly the player's body, the sense of presence and embodiment are maximized (Fig. \ref{fig:roles}a). 

Another common way of interacting with the game is by exploiting tangible interfaces (TIs) \cite{baldo, destefanis} or other physical accessories \cite{villani}. This interaction method proved to be very engaging, especially in case of table-top setups\cite{baldo, robotable2}. In some cases, the robot is used as a TI itself \cite{robotable2}: this approach should be adopted only when that robot is not the only robotic element in the game, otherwise it would be considered by the player just as a fancy controller, reducing the RIF (2/3).

\subsection{Agent's Position and Game Spaces}\label{subsec:dof}
\begin{figure}[t!] 
    \centering
        \includegraphics[width=0.79\columnwidth]{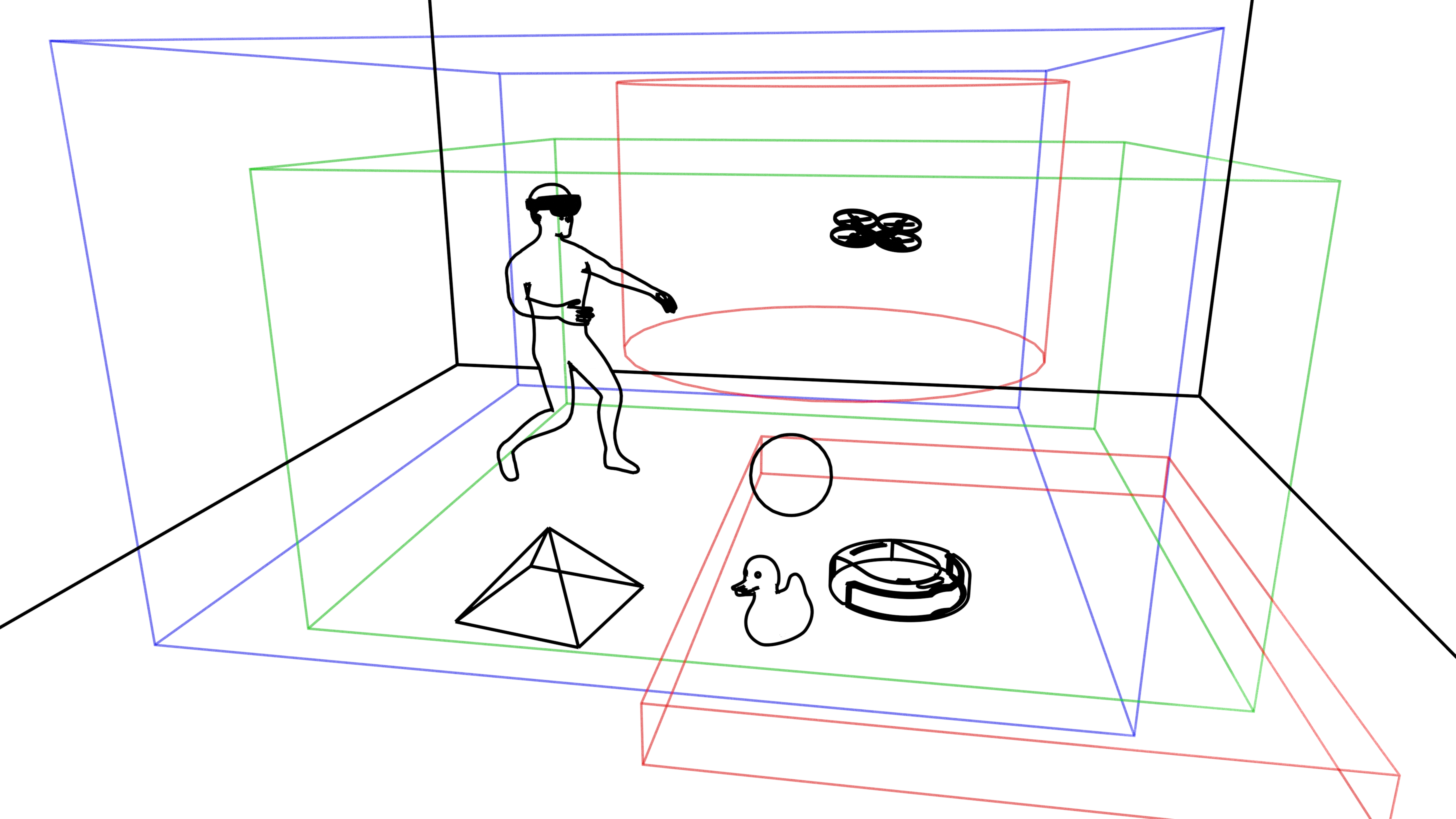}
    \caption{Example of game spaces in a MMRG setup using a headworn MR display, two mobile robots and several physical props (or TIs). Spaces are marked blue for player, green for MR and red for robots. Play space is represented by the room. }
    \label{fig:spaces}
\end{figure}

The player's position in, or w.r.t., the player space, the robot space, the MR space and the play space is a crucial aspect of a MRRG design. The player and robot spaces (or agent spaces) are the physical spaces where the human player(s) and the robot(s), respectively, are allowed/programmed to move during the experience. The MR space (MRS) is the bounding area/volume in which the MR system is able to provide the visual augmentation. For display technologies with a limited field of view (FOV), like head-worn or handheld AR devices, this space is considered as not bounded by the FOV. The play space is the total physical space available for the experience, and its lower-bound is the union of the other three spaces. Examples of different configurations/combinations are given in Fig. \ref{fig:spaces}.

As investigated in \cite{junchao} and \cite{blending}, agent spaces and MRS can match, overlap or be totally disjoint. In IPs that include multiplayer interactions, the organization of player spaces could follow the same approach (delimited space, or shared space). When the player uses the body to interact (though not only in this case), the choice maximizing the RIF is to have a shared space among all the agents or a significant overlap between their spaces. However, this is not sufficient to guarantee a RIF higher than 3. To reach 4 or 5, the relation between the MRS and the agent spaces needs to be considered. In particular, the MRS should not match or contain a whole agent space. In fact, a great RIF boosting factor is represented by the possibility for an agent to make ``incursions'' in that space. The MRS could (actually, should) be manipulated by designers considering the technological boundaries of MRS as an upper-bound. The dynamic manipulation of its location, size and shape during gameplay could be considered too, but it should be carefully designed to provide clear and appropriate feedback to all the players involved in order to avoid confusion.

\subsection{Robot Movement and Interactions}\label{subsec:movstyle}
Even though, as said, MRRGs does not prescribe the use of a specific kind of robots, designers should be aware that robots that can move in space (known as mobile robots) can lead to more engaging experiences compared to robots without this capability. This is due to the fact that the number of degrees of freedom (DOFs), the movement style and the usage of  space have been shown to be key factors to consider with the aim to maximize the RIF.
 
In the following, we consider these capabilities not as upper-bounded by the robot hardware of the actual robot (which, of course, is something to manage), but rather by how they are exploited in the designed game (i.e., as they are seen from the player's perspective): for instance, if the robot is capable of two translational DOFs but in the game, it is moving just along a line, then it will be accounted as 1DOF. Basically, according to the literature, the more the DOFs used in the game, the better the RIF; moreover, translational DOFs are more important than rotational ones \cite{pirone, junchao}.

\begin{figure}[t!] 
    \centering
        \subfloat[]{\includegraphics[width=0.48\columnwidth, height=3.0cm]{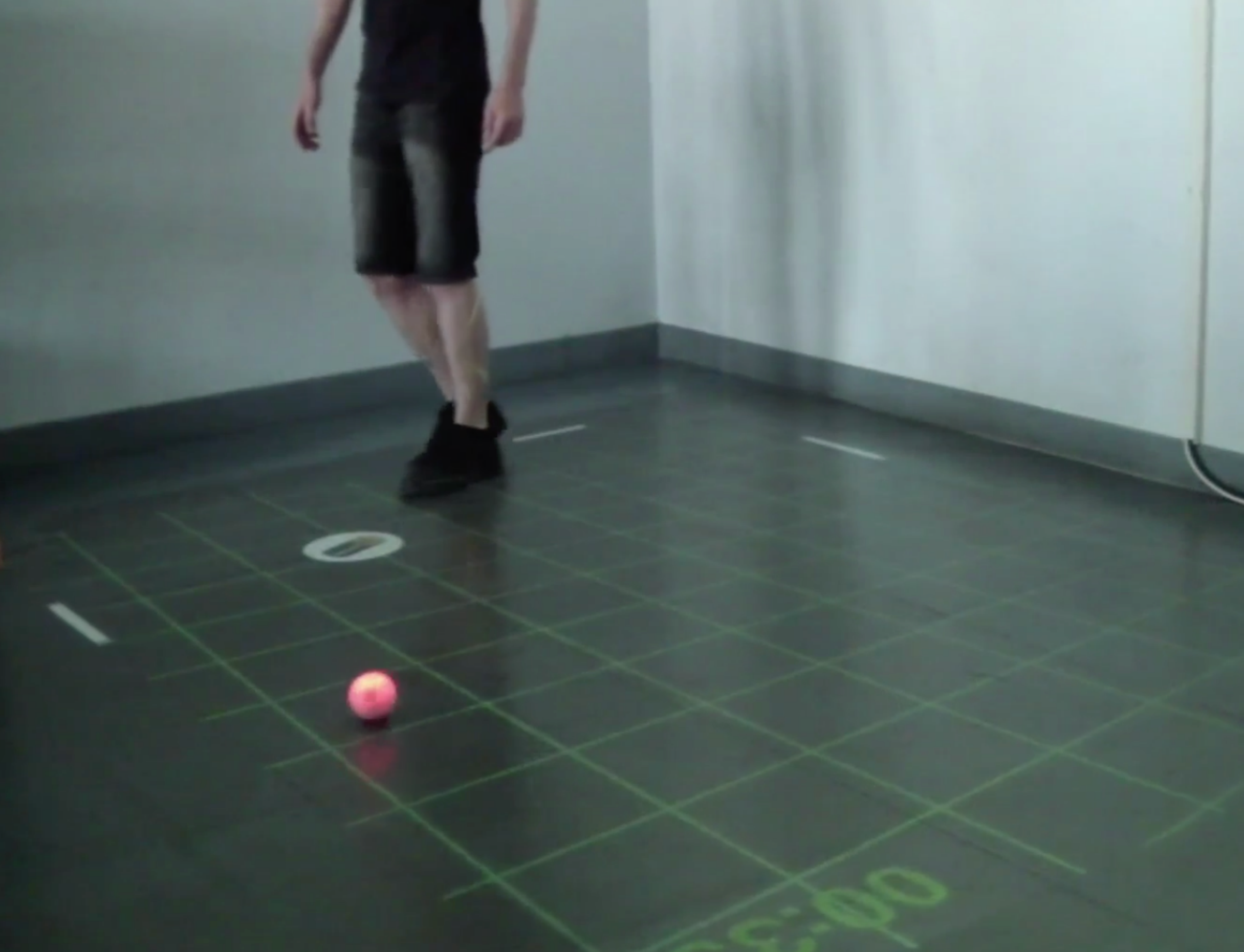}%
        \label{fig:pirg}}
    \hfil
        \subfloat[]{\includegraphics[width=0.48\columnwidth, height=3.0cm]{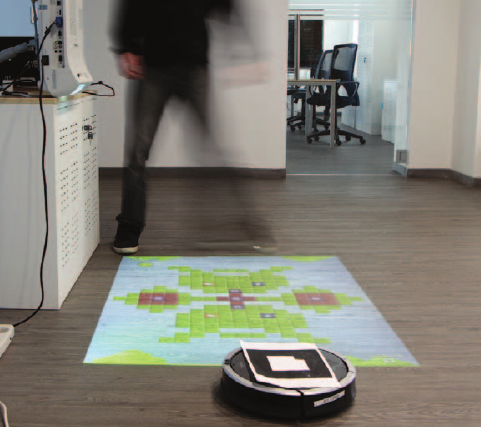}%
        \label{fig:MRRG}}
    \caption{Robot movement: a) MRRG where the ``limitations'' of a spherical robot in terms of motion control where exploited to recreate the natural movement of ball in a balance board game \cite{tld}; b) poor interactivity caused by the use of a slow, vacuum cleaner robot in an Arkanoid-style game \cite{saro}.}
    \label{fig:robot_movement}
\end{figure}

Concerning movement style, this is a broad way to refer to concepts like, e.g., trajectory planning, holonomicity, movement speed and control accuracy as just the final effect that the player experiences during the game \cite{robot_animation}. For instance, the usual practice of making robots follow lines \cite{destefanis}, a curved line helps in improving the RIF more than a broken line, because the robot tends to appear as more lifelike. However,  perfectly following a curved line was shown to lower down the RIF compared to following a broken line in an imperfect, but the smooth, way (for the same reason). Although it may be considered as counter-intuitive, the imperfection of control is actually a boosting factor for the RIF (4/5), because it promotes the sense of life-likeness in the robot \cite{imperfect_control}, and may be difficult or demanding to reproduce faithfully in a VA. Nevertheless, dealing with a robot that is not able to reach, e.g., a target with sufficient accuracy, may not be always possible: in these cases, adding some ``apparent'' entropy without sacrificing safety shall be preferred to perfect control (Fig. \ref{fig:robot_movement}a). Similar considerations hold also for speed, with the further caveat that choices made shall preserve the player’s engagement (Fig. \ref{fig:robot_movement}b).

As said, robot's movements determine its space, which needs to be balanced with robot's size, DOFs and speed. They also determine relations with other spaces, which can affect the RIF. Consider, e.g., a robot space overlapped with the other spaces in a TB game: if the robot uses its exclusive area only in its pause turn, from the GD perspective its space can be seen as collapsed in the intersection among spaces (to be avoided). In fact, robot incursions in other spaces, especially in the MRS, are particularly effective in increasing the RIF (up to 5) if aimed to enable interactions with the environment (e.g., to alter it \cite{incretable}). Interaction can be both with digital and physical elements (unsurprisingly, some toy robots come bundled with TIs \cite{baldo}). Physical interaction between human players and robots should be promoted too, taking care of safety aspects.

Focusing on the robot role, the less sensitive to the above factors is the RK, especially if supported by an IP based on a TB modality: in this case, even a robot with 1DOF and a small area could be good enough for a satisfactory RIF (4) \cite{baldo}. For the BD role, it is suggested to have at least two translational DOFs, and it is acceptable to have a robot space disjoint from the player one. The most demanding role is the ADV one, where all the above cues should be considered, whenever possible. 

\subsection{Spatialized Audio and Voice Interaction}\label{subsec:audio}
\begin{figure}[t!] 
    \centering
        \subfloat[]{\includegraphics[width=0.48\columnwidth, height=3cm]{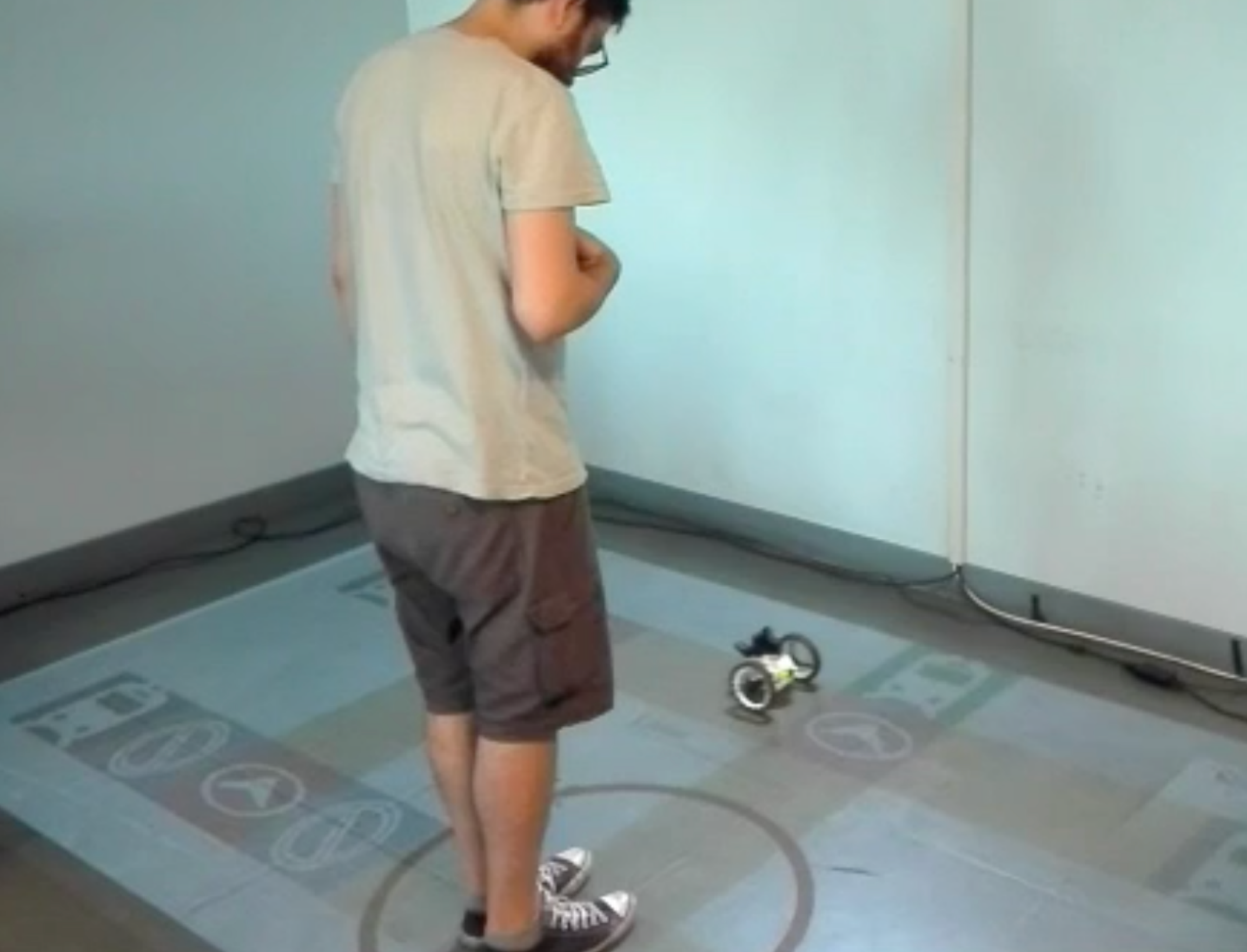}%
        \label{fig:pirg}}
    \hfil
        \subfloat[]{\includegraphics[width=0.48\columnwidth, height=3cm]{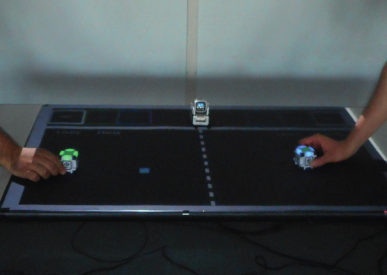}%
        \label{fig:MRRG}}
    \caption{Audio and voice interaction: a) MRRG game used for vestibular disorders rehabilitation, in which the robot is driving the player using sound \cite{pirone}; b) turn-based, Simon-style game in which the robot pronounces color sequences to reproduce \cite{baldo}.}
    \label{fig:audio}
\end{figure}

A more than nice-to-have robot feature to use in a MRRG is the capability to reproduce sounds (Fig. \ref{fig:audio}). In fact, spatialized 3D audio is demonstrated to enhance significantly the immersion and enjoyability in MR experiences. However, there are still technological challenges and setup constraints associated with the use of real-time, immersive audio. Thus, as shown in \cite{pirone}, having a robot that emits sounds while moving in the space can largely increase the RIF (5), since it can be easily localized by the player even when it is out of its FOV. 

Voice interaction was explored as well \cite{baldo, akiko}. For the BD and the RK roles, it could improve the RIF, whereas in the case of the ADV it would not have a clear impact. It is worth stressing that voice interaction must be carefully designed, as it should be entangled to the emotional features of the robot.

\subsection{MR Contents}\label{subsec:mrc}
Besides sounds, another crucial choice to make while designing a MRRG concerns (visual) contents to be delivered to the player via MR augmentation. Providing few augmented contents could improve the RIF, but would dramatically diminish the usefulness of MR. On the other hand, too many augmented contents would be harmful for the RIF.

As a general rule of thumb, the designer should take advantage of the MR setup only (or mostly) to provide the context for the game \cite{baldo}. In principle, no primary game agents and no resources that interact with the player or with the robot should be only digital. Thus, situations in which, e.g., a digital character is used to replace the robot entirely (like in \cite{aracno}) shall be avoided, as they clearly have bad consequences on the RIF (1/2). As said, the use of virtual objects for secondary resources (collectibles, power-ups, etc.) or context agents is bearable. 

A good, negative example could be a pong game \cite{saro}, in which a fast-moving virtual ball is the main agent: player's attention would be mostly directed to that object, and the role of a robot in a possible MRRG design  would necessarily be minor. Another negative example could be a racing game with a digital-only circuit: this design would ruin the RIF (1/2), as the circuit is a game element that is meant to primarily interact with phygital agents and is supposed to be able to constrain their movement.

A proper use of (visual) augmentation is the provisioning of general system feedback and of special visual effects. Good examples could be, e.g., the use of visual MR contents to enhance the emotional capabilities of a robot \cite{farting}, or to give a particular connotation or identity to the game agents game (like health, available accessories, etc.).

\section{Technologies}

Most of the games mentioned above were created with Unity, a game engine that can be used to manage both the application logic and the AR visualization. Headset-based AR contents could be dealt with using vendor-provided Software Development Kits (SDKs), like the Microsoft MR toolkit, which are easy to integrate in Unity. When external (robot, hand or body) tracking is required, either software methods, e.g., relying on RGB or depth images (which can be processed using open source libraries like OpenCV, or commercial frameworks like the Microsoft Azure Kinect DK or the Intel RealSense SDK) or hardware-based solutions (e.g., leveraging Valve's Lighthouse trackers together with commercial or open solutions like SteamVR or OpenXR) could be adopted. Robot control is usually achieved through vendor-provided SDKs, often declined as ROS (Robot Operating System) modules.

\section{Conclusions and Future works} \label{sec:limit}

In this paper, we introduced a formal definition for MRRGs and discussed key challenges faced in creating these game experiences. 
We then provided a set of multifaceted GD guidelines to consider for dealing with the above challenges, focusing in particular on approaches aimed to maximize the central role played by the robotic components. 

The approach adopted is similar to those pursued for PIRGs\cite{pirg}, with guidelines simultaneously proposed and validated through dedicated user studies. Hence, in most of the cases, design hints are supported by statistical and factual pieces of evidence, which are reported in the referenced literature works. However, like for PIRGs, some of the considerations are based on knowledge pertaining every-day practice that characterized our exploration of this field in the last five years. This knowledge includes findings based, e.g., on transitional experiments or feedback collected through interviews with players involved in the user studies, which went unreported in the published papers. 

Hence, in order to validate these guidelines, further work by researchers active in the field will be required. For instance, although mobile robots are expected to provide more engaging experiences, tests will have to be performed with other robots like, e.g., robotic arms. Effort should be also devoted to investigate complex multiplayers, multilateral IPs, involving non-homogeneous robots and robot swarms, as well as variations on robot controls and HRI methods that can enable safe interaction of fast-moving robots and players. From the perspective of CE, it will be particularly important to study other types of AR-based visualizations and their impact on RIF, as well as the acceptance of possible, future scenarios in which service robots carry out the task(s) they have been designed for in MR gamified environments. 

Despite these limitations, reported findings represent a synthesis of the most relevant experimental studies carried out so far by the research community in this field, which are expected to allow developers to create ever more enjoyable experiences. Hopefully, motivated by this work other researchers will join our ongoing effort and contribute at widening this set of guidelines and at strengthening their (combined) validation in both objective and subjective terms. Results could then foster further developments in the robotic gaming domain and push the market of related CE technologies.

\bibliographystyle{IEEEtran}
\bibliography{paper_final}
\vspace{-1.5cm}
\begin{IEEEbiography}{Filippo Gabriele Prattic\`o} is a Ph.D. student at Politecnico di Torino. His interests pertain VR, AR, HMI, serious games and UX design. Contact him at \href{filippogabriele.prattico@polito.it}{filippogabriele.prattico@polito.it}
\end{IEEEbiography}
\vspace{-1.5cm}
\begin{IEEEbiography}{Fabrizio Lamberti}
is a full professor at Politecnico di Torino. His interests are in the areas of computer graphics, HMI, and intelligent systems. 
He serves as Associate Editor for several IEEE Transactions and for IEEE Consumer Electronics Magazine. 
Contact him at \href{fabrizio.lamberti@polito.it}{fabrizio.lamberti@polito.it}
\end{IEEEbiography}
\end{document}